\documentclass{ws-procs10x7}
\usepackage{balance}

\newcolumntype{d}[1]{D{.}{.}{#1}}

\def\Journal#1#2#3#4{{\it #1} {\bf #2}, #3 (#4)}

\def\JPG{{\em J. Phys.} G}

\def\EPJC{{\em Eur. Phys. J.} C}

\def\PRT{\em Phys. Rept.}

\def\etal{{\it et al.}}

\def\NPPS{\em Nucl.\ Phys.\ Proc.\ Suppl.}

\newcommand{\epem}              {\ensuremath{\mathrm{e^+e^-}}}
\newcommand{\as}                {\ensuremath{\alpha_\mathrm{S}}}
\newcommand{\az}               {\ensuremath{\alpha_0}}
\newcommand{\asmz}              {\ensuremath{\alpha_\mathrm{S}(M_{\mathrm{Z^0}})}}
\newcommand{\qqbar}     {\ensuremath{\mathrm{q\bar{q}}}}
\newcommand{\bbbar}     {\ensuremath{\mathrm{b\bar{b}}}}

\makeindex
\begin{document}

\title{Measurement of \as\ in \epem\ collisions at LEP and JADE}

\author{J. Schieck$^*$}

\address{Max-Planck-Institut f\"ur Physik, F\"ohringer Ring 6, 80805 M\"unchen
 \\$^*$E-mail: schieck@mppmu.mpg.de}


\twocolumn[\maketitle\abstract{
  Data from \epem\ annihilation into hadrons collected by the JADE, the L3 
  and the OPAL experiment at centre-of-mass energies between 14~GeV and 
  209~GeV are used to determine the strong coupling \as. Observables in 
  leading order sensitive to \as\ as well as $\as^{2}$ are used.
  The evolution of \as\ with respect
  to the centre-of-mass energy as predicted by QCD is studied and 
  confirmed with high precision. All measurements of \as\ 
  are consistent with the current world average.
}
]
\section{Introduction}
The clean environment of \epem\ annihilation allows to access the 
strong coupling \as, the free parameter of Quantum Chromodynamics (QCD),
very well. The strong coupling \as\ reflects the probability of the 
radiation of gluons, the vector gauge boson of QCD. 
General event topologies can be studied or the number of 
radiated hard gluons, identified by jets, can be counted. 
To numerically evaluate the 
strength of the strong coupling \as\ the measured observables have to be
compared to theoretical predictions. These theoretical calculations predict
the radiation of gluons with \as\ being the only free parameter. 
Here the strong coupling \as\ is studied in the \epem\ energy range 
between 14 GeV and 209 GeV. The energy evolution of the strong coupling \as\ is
studied and compared to the prediction of QCD. A detailed summary of
the results can be found at~\cite{l3,opal_jet,opal_4jet,jade} and
references therein.
\section{Data Sample}
Data taken with the JADE experiment cover energy points in
the energy range between 14 and 44 GeV. At each energy point 
between 1k and 20k multihadronic events are selected. 
The L3 collaboration uses events with hard inital or final state
radiation (radiative events) to access
the energy ranges below the centre-of-mass energy of the colliding beams 
in the range between 30 and 86 GeV. At every point between 1k and 3k events are 
selected. For the determination of the strong coupling 
\as\ factorization between the gluon and the photon
production is assumed. Theoretically the energy scale of the process 
is not well defined~\cite{salam}.
The largest dataset selected at L3 and OPAL is at a centre-of-mass 
energy of 91 GeV with more than 100k events.  Above 91 GeV events are selected 
with 500-5k events in several energy points up to 209 GeV. \\
Different sources of background are considered at the various energy points. At
JADE  \bbbar\ events are subtracted since the electroweak decay of B-mesons
fakes events with hard gluon radiation and lead to a bias in the measurement 
of \as. Above centre-of-mass energies of 
130~GeV the large number of radiative events are rejected as background,
since they lower the effective centre-of-mass energy. At energies above 
160~GeV hadronic decays of W-pair events are subtracted.
At the centre-of-mass energy of 91 GeV no background is considered.
\section{Event-Shape Observables}
\begin{figure}[h]
\centerline{\psfig{file=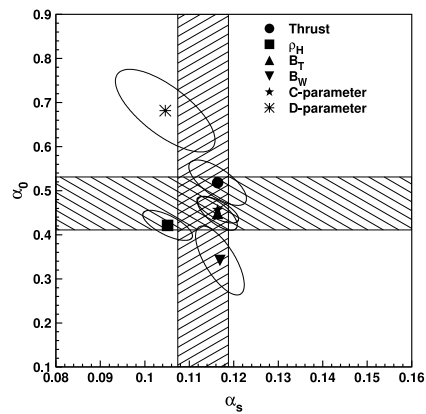,width=2.2in}}
\caption{The values of \as\ and $\alpha_{0}$ from fits of the
power correction ansatz to the first moment of the six event-shape
observables. The ellipses represents $39\%$ two-dimensional confidence
intervals including both statistical and systematic uncertainties. The band
represent unweighted averages of the \as\ and $\alpha_{0}$ including both
statistical and systematic uncertainties.}
\label{powercorr}
\end{figure}

The radiation of hard gluons in the \qqbar\ final state alters the topology
of the event. Events without hard gluon radiation look more pencil like, while 
the radiation of hard gluons make the event look more spherical. The size
of the strong coupling \as\ therefore has an impact on the event
topology. This impact on the topology can be quantitatively described
by so called event-shape observables, which are calculated using the
charged particle tracks and neutral clusters of the event. The L3 Collaboration uses 
event-shape observables like Thrust, C-Parameter, the
jet broadening variables $B_{T}$ and $B_{W}$, the jet resolution parameters $y_{23}$
and the heavy jet mass to determine the strong coupling \as. All these 
event-shape observables are in leading order 
proportional to \as. In addition the L3 and the OPAL collaboration study the D-parameter 
and Thrust-Minor, which are proportional to $\as^{2}$ in leading order. For the measurement 
of \as\ using event-shape observables events with a centre-of-mass energy
between 91 and 209 GeV and radiative events with an effective 
centre-of-mass energy between 30 and 86 GeV are used.
QCD calculations predict the distribution or the mean value of 
event-shape observables as a function of the strong coupling \as.
The predictions are calculated at parton level only and a correction
for hadronization effects has to be taken into account. \\
The first moments of the event observables $\langle F \rangle = 
\int F \frac{1}{\sigma} \frac{d\sigma}{dF}dF$, with F being the 
event-shape observable and $\sigma$ being the cross-section, are used to 
determine the strong coupling \as. This approach allows to sample the
full region of available phase-space. The QCD predictions are available in 
next-to-leading order (NLO). The hadronization effects are described by
power corrections, which predict the changes of the event-shape observable 
due to hadronization with a single parameter \az.
The results of \as\ and \az\ are shown in Fig.~\ref{powercorr}.
The unweighted average of the six measurements is:
\begin{eqnarray}
\asmz = 0.1126   \pm 0.0045\pm0.0039 \nonumber \\
\alpha_{0}=0.47\pm 0.054\pm0.024, \nonumber
\end{eqnarray}
where the first uncertainty corresponds to the statistical and the second
to the theoretical one.
The confidence level for a common $\alpha_{0}$ is $3\%$, if the systematic 
uncertainties are treated as uncorrelated. Theoretical 
uncertainties from the power correction approach are not considered.\\
In a further measurement fits to the event-shape distributions 
are performed. In these fits only parts of the available phase-space is used 
to determine \as. The theoretical predications are available
in NLO combined with resummed calculations (NLLA), leading to an 
improved description of the data.
The hadronization is evaluated using Pythia Monte Carlo and Herwig and 
Ariadne as a systematic uncertainty. The combined result is
\begin{eqnarray}
\asmz = 0.1227   \pm 0.0012\pm0.0058, \nonumber 
\end{eqnarray}
where the first uncertainty corresponds to the experimental one and the second
to the theoretical one. The theoretical uncertainties contain errors
from missing higher order terms in the calculation and errors associated
to the hadronization correction. 
The combination of the \as\ measurements using event-shape observables at the
varies centre-of-mass energies assumes the validity of QCD, in particular
the energy evolution of \as\ with energy. The combination assuming
the energy evolution of \as\ according to QCD results a $\chi^{2}$ value 
of $17.9/15$. A fit
with a constant value of \as\ returns a $\chi^{2}$ value of $51.7/15$. \\
Measurements of \as\ using event-shape distributions of event-shape observables proportional
to $\as^{2}$ in leading order, like the D-Parameter and Thrust minor
are performed as well. Only NLO predictions are available, leading to an 
increased theoretical uncertainty.
\section{Measurements of \as\ using Jet Rates}
\begin{figure}[h]
\centerline{\psfig{file=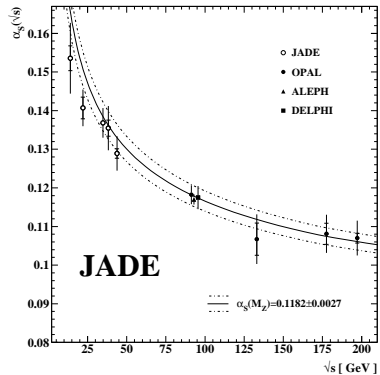,width=2.2in}}
\caption{The values of \as\ at the various energy points. The
errors show the statistical (inner part) and the total error.
The full and dash-dotted lines indicate the world average
value of \as\ with error.}
\label{alphas}
\end{figure}

Another way to determine the strong coupling \as\ is to measure the number
of selected jets. For this all particles have to be clustered according to
a certain jet-finder scheme. The OPAL collaboration applies two different schemes, the 
Durham scheme and the Cambridge scheme, both with $y_{cut}$ as a free 
parameter. The average jet-rate $\langle N(y_{cut})\rangle =
 \frac{1}{\sigma_{\mathrm{tot}}} \sum_{n} n \sigma_{n}(y_{cut})$ and the
differential 2-jet rate $y_{23}$ is used, both being sensitive 
to \as\ in leading order. The theoretical prediction applied here are 
combined NLO+NLLA calculations. Hadronization correction are determined
with Monte Carlo. The combined value of both jet-finder
schemes applied to the energy range between 91 and 209 GeV is
\begin{eqnarray}
\asmz=0.1177\pm0.0006\pm0.0012 \nonumber \\ 
\pm0.0010\pm0.0032,\nonumber
\end{eqnarray}
with the first uncertainty being the statistical one, the second the 
experimental one, the third due to the hadronization and the last the 
theoretical one. \\
The OPAL and the JADE collaboration uses the number of selected four-jet
events to determine the strong coupling \as. The theoretical predictions
are NLO combined with NLLA calculations. The leading order prediction
is proportional to $\as^{2}$ and therefore the theoretical uncertainty due to higher order 
missing terms is proportional to $\as^{3}$. A decreased theoretical
uncertainty compared to leading order \as\ observables  is expected. 
The combined value of \as\ determined by OPAL using the Durham jet-finder scheme and 
data between 91 and 209 GeV is
\begin{eqnarray}
\asmz=0.1182\pm0.0003\pm0.0015 \nonumber \\ \pm0.0011\pm0.0012\pm0.0013, \nonumber
\end{eqnarray}
with the first uncertainty being the statistical one, the second the 
experimental one, the third due to the hadronization, the fourth due to higher
order missing terms and the last due to effects originating from the b-quark
mass. \\
A similar measurement is performed by the JADE collaboration also using the Durham
jet-finder and data between 22 and 44 GeV. The combination 
returns a value of
\begin{eqnarray}
\asmz=0.1159\pm0.0004\pm0.0012 \nonumber \\ \pm0.0024\pm0.0007, \nonumber
\end{eqnarray}
with the first uncertainty being the statistical one, the second the 
experimental one, the third due to the hadronization and the last the 
theoretical one. Both \as\ measurements have a similar precision, 
dominated by systematic uncertainties.
The theoretical uncertainty due to higher order missing terms is decreased 
compared to measurements of \as\ using observables sensitive in leading order 
to \as. Fig.~\ref{alphas} shows the 
result and the variation of \as\ with respect to the centre-of-mass
energy.
In the case of the \as\ measurements using four-jet events two alternative 
methods to evaluate the theoretical uncertainty are studied.
First, both, the renormalization scale factor and the strong coupling \as\
are varied within the fit. Second, the renormalization scale factor is determined
as the one with the least sensitivity with respect to \as. 
Both alternative methods return values of \as\ which are well within
the theoretical uncertainty of the default method.
The natural choice of the renormalization scale factor 
is close to the renormalization scale factor having the
least sensitivity to the renormalization scale factor.
This leads to smaller theoretical uncertainties
determined by varying the scale factor, compared 
to fits to event-shape observables sensitive to \as\ in
leading order. \\
Measurements at OPAL and JADE alone return no significant
proof for the running of \as. However, a combined fit to the
OPAL and JADE data points confirms the running with high significance
with a $\chi^{2}$ value of 12.0/18, compared to 149.5/18 assuming a constant 
value of \as.
\section{Summary}
\begin{figure}[h]
\centerline{\psfig{file=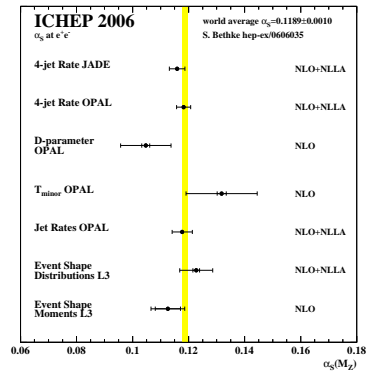,width=2.2in}}
\caption{A summary of all measurements of \as\ discussed
in this article. All measurements are consistent 
with the current world average.}
\label{summary}
\end{figure}

Studies of the strong coupling using event-shape observables and 
jet-rates in \epem\ are presented. All 
measurements return values of \as\ consistent with 
the world average~\cite{bethke}. A summary of all measurements
is shown in Fig.~\ref{summary}. The determination
of \as\ using the four-jet rate leads to a decreased
theoretical uncertainty. The combination of 
data taken at LEP experiments and the JADE experiment
confirm the running of \as\ with high significance.

\end{document}